\documentclass[11pt]{article}

\usepackage{mathtools}
\usepackage{graphicx}
\usepackage[author-year]{amsrefs}

\title{\bf Cross validation residuals for generalised least squares and other correlated data models}
\author{Ingrid Annette Baade\\
        Grid Statistical Services\\
        \texttt{ingridbaade1@gmail.com}}
\date{}  

\begin{document}
\maketitle
\begin{abstract}
Cross validation residuals are well known for the ordinary least squares model. Here leave-M-out cross validation is extended to generalised least squares. The relationship between cross validation residuals and Cook's distance is demonstrated, in terms of an approximation to the difference in the generalised residual sum of squares for a model fit to all the data (training and test) and a model fit to a reduced dataset (training data only). For generalised least squares, as for ordinary least squares, there is no need to refit the model to reduced size datasets as all the values for K fold cross validation are available after fitting the model to all the data. 
\end{abstract}

\section{Introduction}

Cross validation is an important method, commonly used in machine learning and other statistical analyses (see for example, \citeauthory{Hastie09}). Ordinary least squares (OLS) cross validation methods can be adapted for generalized linear models, but there are many models in common use that do not assume uncorrelated errors. In this paper we review the derivation of leave M out cross validation statistics and associated quantities for OLS and then develop these quantities for generalised least squares (GLS). 

\section{Cross validation residuals for ordinary least squares}

Under an OLS model with known variance $\sigma^{2}I$, let 
\begin{align*}
r^{*_{(M)}} = Y-X \hat{\beta}_{(M)} = \begin{bmatrix} Y_{(M)}-X_{(M)} \hat{\beta}_{(M)} \\ Y_{M}-X_{M} \hat{\beta}_{(M)} \end{bmatrix} = \begin{bmatrix} r^{*_{(M)}}_{(M)} \\r^{*_{(M)}}_{M} \end{bmatrix}, 
\end{align*}
where $r^{*_{(M)}}_{(M)}$ are the residuals from fitting the OLS model to all the data except a group of size $m$ of the observations indexed by $M$. The superscript $\phantom{}^{*_{(M)}}$ is a label denoting this model; the brackets around the $M$ can be read as ``without the group M". The other component of $r^{*_{(M)}}$ is $r^{*_{(M)}}_{M}$, the residuals predicted for the $M$ cases not used in the model. For OLS the collection $r^{*_{(M)}}_{M}$ of $m$ quantities is also called the leave-M-out cross validation (LMOCV) residual. 
We have
\begin{align*}
r^{*_{(M)}}_{M} &= Y_{M} - X_{M}\hat{\beta}_{(M)} \\
  &= Y_{M} - X_{M}\hat{\beta}+ X_{M}\hat{\beta}- X_{M}\hat{\beta}_{(M)} \\
  &= r_{M} + H_{M} (I_{m} - H_{M})^{-1} r_{M}  \\
  &= (I_{m} - H_{M})^{-1}r_{M}.
\end{align*}

The residual sum of squares from the model without $M$ cases is
\begin{align*}
r^{*_{(M)}T}_{(M)} r^{*_{(M)}}_{(M)}
& = (Y_{(M)}-X_{(M)} \hat{\beta}_{(M)})^{T}(Y_{(M)}-X_{(M)} \hat{\beta}_{(M)}) \\
&= (Y-X \hat{\beta}_{(M)})^{T} \begin{bmatrix} I_{n} &- \begin{bmatrix} 0 & 0 \\ 0 & I_{m} 
                                                        \end{bmatrix}    
                               \end{bmatrix} 
(Y-X\hat{\beta}_{(M)}) \\
&= \begin{multlined}[t](Y-X\hat{\beta}_{(M)})^{T}(Y-X\hat{\beta}_{(M)})  \\
     - (Y_{M}-X_{M}\hat{\beta}_{(M)})^{T}(Y_{M}-X_{M}\hat{\beta}_{(M)})  
   \end{multlined}\\
&= \begin{multlined}[t](Y-X\hat{\beta})^{T}(Y-X\hat{\beta})  \\
              + r_{M}^{T}(I_{m}-H_{M})^{-1}H_{M}(I_{m}-H_{M})^{-1}r_{M} \\ 
              - r_{M}^{T}(I_{m}-H_{M})^{-1}(I_{m}-H_{M})^{-1}r_{M}. \\
   \end{multlined}
\end{align*}
Details of the algebra are in the appendix, and also see \citeauthory{Baade98}. The second term on the right hand side in the last line is a multiple of Cook's distance for cases $M$, and if $m=1$, the third term is the sum of squares of the LOOCV residuals.  Simplifying the right hand side of this expression gives
\begin{align}
r^{*_{(M)}T}_{(M)} r^{*_{(M)}}_{(M)}
&= r^{T}r - r_{M}^{T}(I_{m}-H_{M})^{-1}r_{M}.
\label{eqn:olsSS1}
\end{align}
In this paper we refer to the difference between the ``square" of the cross validation residual and the Cook's distance type quantity, as the ``squared residual difference" SRD:
\begin{align*}
\textnormal{SRD} &= \textnormal{LMOCV}^2 - k \times \textnormal{CD} \\
&= r_{M}^{T}(I_{m}-H_{M})^{-1}r_{M}. 
\end{align*}

When $\sigma^{2}$ is not known, it is estimated by $\hat{\sigma}^{2} = \frac{1}{n-p}r^{T}r$ under restricted maximum likelihood estimation. For a model based on a reduced size dataset of $n-m$ cases, $\sigma^{2}_{*(M)}$ is estimated by $\hat{\sigma}^{2}_{*(M)} = \frac{1}{n-p-m}r^{*_{(M)}T}_{(M)}r^{*_{(M)}}_{(M)}$. Equation~\ref{eqn:olsSS1} becomes
\begin{align*}
(n-p-m)\hat{\sigma}^{2}_{*(M)}
&= (n-p)\hat{\sigma}^{2} - r_{M}^{T}(I_{m}-H_{M})^{-1}r_{M}. \\
\end{align*}

\section{Extension of cross validation residuals to generalised least squares}

In this paper, for GLS, we consider models where $\textnormal{var}(Y) = V = \sigma^2 \Sigma$, where $\Sigma$ is a correlation matrix. (Future work will consider more general variance matrices, incorporating heteroscedasticity.) Assume initially that the parameters $\rho$ on which $\Sigma$ depends, are known and do not change when a model is fitted to a reduced size dataset. The (generalised) residual sum of squares can be broken down in a similar way to the OLS case:
\begin{align*}
& r^{*_{(M)}T}_{(M)} (\Sigma_{(M)})^{-1}r^{*_{(M)}}_{(M)} \\
& \phantom{\hat{\beta}_{(M)}} = (Y_{(M)}-X_{(M)} \hat{\beta}_{(M)})^{T}(\Sigma_{(M)})^{-1}(Y_{(M)}-X_{(M)} \hat{\beta}_{(M)}) \\
& \phantom{\hat{\beta}_{(M)}} = (Y-X \hat{\beta}_{(M)})^{T} \left[ \Sigma^{-1} - \Sigma^{M \textnormal{cols}}(\Sigma^{M})^{-1} \Sigma^{M \textnormal{rows}}\right] (Y-X\hat{\beta}_{(M)}). 
\end{align*}
Define $\widetilde{Y} = \Sigma^{-1}Y$, $\widetilde{X} = \Sigma^{-1}X$, $\widetilde{r} = \Sigma^{-1}r$ and $\widetilde{H} = \Sigma^{-1}H \Sigma^{-1}$, where $H = X(X^{T}\Sigma^{-1}X)^{-1}X^{T}$. Then
\begin{align} 
& r^{*_{(M)}T}_{(M)} (\Sigma_{(M)})^{-1}r^{*_{(M)}}_{(M)} \nonumber \\
& \phantom{\hat{\beta}_{(M)}}   = (Y-X\hat{\beta})^{T}\Sigma^{-1}(Y-X\hat{\beta}) + \widetilde{r}_{M}^{T}(\Sigma^{M}-\widetilde{H}_{M})^{-1}\widetilde{H}_{M}(\Sigma^{M}-\widetilde{H}_{M})^{-1}\widetilde{r}_{M} \nonumber \\ 
& \phantom{\hat{\beta}_{(M)}\hat{\beta}_{(M)} }    - \widetilde{r}_{M}^{T}(\Sigma^{M}-\widetilde{H}_{M})^{-1}\Sigma^{M}(\Sigma^{M}-\widetilde{H}_{M})^{-1}\widetilde{r}_{M}.
\label{eqn:glsSS1} 
\end{align}
Details of the derivation are in the appendix. $\sigma^2$ is estimated by $\frac{1}{n-p}r^{T}\Sigma^{-1}r$ and $\sigma^{2}_{*(M)}$ is estimated by $\frac{1}{n-p-m}r^{*_{(M)}T}_{(M)} (\Sigma_{(M)})^{-1}r^{*_{(M)}}_{(M)}$. Equation~\ref{eqn:glsSS1} becomes 
\begin{align}
(n-p-m)\hat{\sigma}^{2}_{*(M)}
&= (n-p)\hat{\sigma}^{2} - \widetilde{r}_{M}^{T}(\Sigma^{M}-\widetilde{H}_{M})^{-1}\widetilde{r}_{M}.
\label{eqn:glsSS2}
\end{align}
The expression
\begin{align*}
\widetilde{r}_{M}^{T}(\Sigma^{M}-\widetilde{H}_{M})^{-1}\widetilde{H}_{M}(\Sigma^{M}-\widetilde{H}_{M})^{-1}\widetilde{r}_{M}
\end{align*}
in equation~\ref{eqn:glsSS1} is a multiple of Cook's distance for GLS \cites{Baade98,Baade00}. 

In many applications, the variance matrix is a function of a scaling parameter $\sigma^{2}$ and one other parameter $\rho$. For example, the CAR1 continuous time autoregressive covariance structure can be used for multiple observations taken from the same person over time. Times may be unequally spaced so the covariance between two measurements on the same individual can be modelled as $\sigma^{2} \rho^{|t_{1} - t_{2}|}$. The resulting variance matrix is block diagonal with blocks corresponding to the measurements over time on a single individual. The inverse of the variance matrix is therefore also block diagonal and if the correlation structure is CAR1, the inverse variance matrix blocks are tridiagonal.  

We defined $\widetilde{Y} $ as $\Sigma^{-1}Y$ but equivalently we could consider  $Y^{*} = S^{1/2}\Sigma^{-1}Y$, where $S$ is the diagonal matrix with elements $1/\Sigma^{ii}$ on the diagonal. The variance of $Y^{*}$ is then 
$\sigma^2 S^{1/2}\Sigma^{-1}S^{1/2}$, where $S^{1/2}\Sigma^{-1}S^{1/2}$ is the partial correlation matrix of $Y$.   
With this notation, equation~\ref{eqn:glsSS2} becomes
\begin{align}
(n-p-m)\hat{\sigma}^{2}_{*(M)}
&= (n-p)\hat{\sigma}^{2} - {r}_{M}^{*T}(C^{M}-{H}^{*}_{M})^{-1}r_{M}^{*}
\label{eqn:glsSS3}
\end{align}
where $C= S^{1/2}\Sigma^{-1}S^{1/2}$ has 1s on the diagonal and $H^{*} = S^{1/2}\Sigma^{-1}H\Sigma^{-1}S^{1/2}$. In particular, for $m=1$, the squared residual difference term is $\frac{r_{i}^{*2}}{1-H_{i}^{*}}$, being the difference between the square of the LOOCV residual $\frac{r_{i}^{*2}}{(1-H_{i}^{*})^2}$ and the Cook's distance type quantity $\frac{H_{i}^{*}r_{i}^{*2}}{(1-H_{i}^{*})^2}$. 
 
Another transformation that might be of interest is $Y^\dagger = S \Sigma^{-1}Y$. Then residuals $r_{i}^\dagger$ can be written as $r_{i} + \sum_{j \neq i} \frac{\Sigma^{ij}}{\Sigma^{ii}}r_{j}$. That is, the $r_{i}^{\dagger}$ are the $r_{i}$ adjusted for correlation with other residuals.   

$Y$ and $\tilde{Y} = \Sigma^{-1} Y$ form a dual basis \cites{156812} for if we define an inner product of $Y_{i}$ and $Y_{j}$ as $\langle Y_{i}, Y_{j} \rangle = \Sigma_{ij}$ then as $\tilde{Y}_{i} = \sum_{k=1}^{n} \Sigma^{ik}Y_{k}$, 
$\langle \tilde{Y}_{i}, Y_{j} \rangle = \sum_{k=1}^{n} \Sigma^{ik} \langle Y_{k}, Y_{j} \rangle  = \sum_{k=1}^{n} \Sigma^{ik} \Sigma_{kj} = \delta_{ij}$, the Kronecker delta.

Through an example, we examine how well equation~{\ref{eqn:glsSS2} holds when $\rho$ is not regarded as constant. 

\section{Example}

\begin{figure}
  \includegraphics[width=\linewidth]{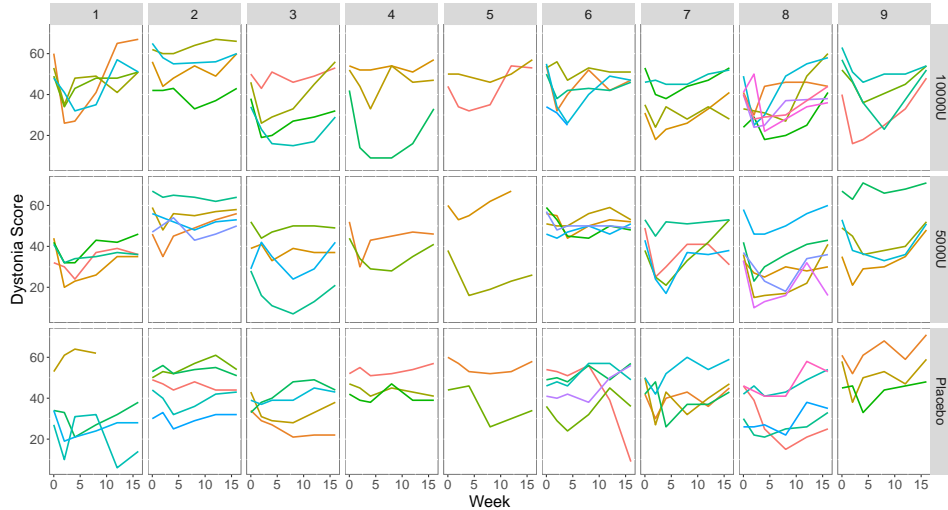}
  \caption{Adapted from \citeauthory{Harrell15}. Scores for 108 individuals at nine study sites. The score at time 0 is one of the explanatory variables. The other scores, at weeks 2, 4, 8, 12, and 16, are modelled with an AR(1) correlation structure.}
\label{fig:data} 
\end{figure}
We fit an AR(1) model to the cervical dystonia data originally recorded in \citeauthory{Davis02} and included with r package ``rms" \cite{rms}. Specifically, ``Gls" from the ``rms" package, based on ``gls" from the ``nlme" package \cite{nlme} was used to fit a generalised least squares model to 522 observations. The dimension of the explanatory variable matrix was 522 by $p=18$.  Patients with cervical dystonia were scored on a rating scale of impairment from cervical dystonia on five occasions (at $t$ = 2, 4, 8, 12 and 16 weeks) after initial measurement and randomisation at $t$ = 0 into three treatment groups: placebo, 5000 units botulinum or 10000 units botulinum. Nine study sites took part. Of the 109 patients, 108 returned on at least one occasion. Ninety four patients were reassessed on all five occasions and a further 11 were reassessed on four occasions.  A plot of the data, figure \ref{fig:data},  shows that there may have been an initial improvement in the non placebo groups but the scores tend to revert to their higher level by the end of the study. There is not much scope for observations to have unusual leverage because the only explanatory variable value an observation can have that is not shared with multiple other individuals is age. The median age is 56 and ages range from 26 to 83. We follow the analysis in \citeauthory{Harrell15} and fit a CAR1 continuous time autoregressive correlation structure. 

\begin{figure}
  \includegraphics[width=\linewidth]{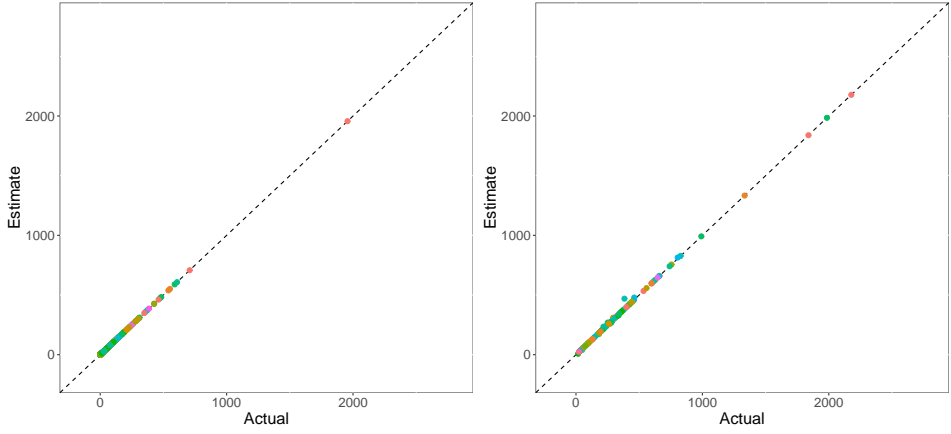}
  \caption{Estimated, $\widetilde{r}_{M}^{T}(\Sigma^{M}-\widetilde{H}_{M})^{-1}\widetilde{r}_{M}$, versus actual, $(n-p)\hat{\sigma}^{2} - (n-p-m)\hat{\sigma}^{2}_{*(M)}$, squared residual difference (SRD) for Leave-One-Out (left) and Leave-M-Out (right) diagnostics. These plots are for OLS so $\widetilde{r}_{M}^{T}(\Sigma^{M}-\widetilde{H}_{M})^{-1}\widetilde{r}_{M} = r_{M}^{T}(I_{m}-{H}_{M})^{-1}r_{M}$.}
\label{fig:OLS} 
\end{figure} 

We begin by fitting an OLS model to the data, and examining the actual differences in the residual sums of squares  calculated from refitting the model without each observation, or group of observations, in turn, $(n-p)\hat{\sigma}^{2} - (n-p-1)\hat{\sigma}^{2}_{*(M)}$, and the estimate arising from simply fitting the model once to all the data, $\widetilde{r}_{M}^{T}(\Sigma^{M}-\widetilde{H}_{M})^{-1}\widetilde{r}_{M} $. For OLS the estimate should be exact, up to the accuracy of the model fitting procedure. The plot on the left in figure~\ref{fig:OLS} shows that for removing only one observation at a time, the actual change in the residual sum of squares agrees closely with the estimate. The observation with the very high SRD corresponds to the last observation of an individual from site 6 in the placebo group; the person's score plummets at week 16, resulting in a large negative residual. 

When all observations for an individual are removed together, the actual change in the residual sum of squares has, for one individual, deviated from the estimated SRD value, as shown on the right in figure~\ref{fig:OLS}. The individual was in the placebo group at site 2 and his scores can be seen in figure~\ref{fig:data} with the second lowest baseline score for this group. The Cook's distance component of the SRD for the individual is large both in absolute value and as a proportion of the SRD. Large Cook's distances come about through high leverage, high residuals, or both. In this case, the observations for this individual have high leverage and high joint leverage. The person is the youngest male in the study. Without his data, the estimate of $\beta$ changes substantially. The location where the restricted likelihood is maximised, ie $\hat{\beta}_{(M)}$, may have been ascertained with less accuracy than the estimates of $\beta$ when other individuals are left out, leading to a less accurate estimate of $\sigma^{2}_{*(M)}$.

Next we fit the AR(1) model to the data, comparing estimated and actual values of the squared residual difference. We examine the effect of the changing correlation parameter value $\rho$. 

\begin{figure}
  \includegraphics[width=\linewidth]{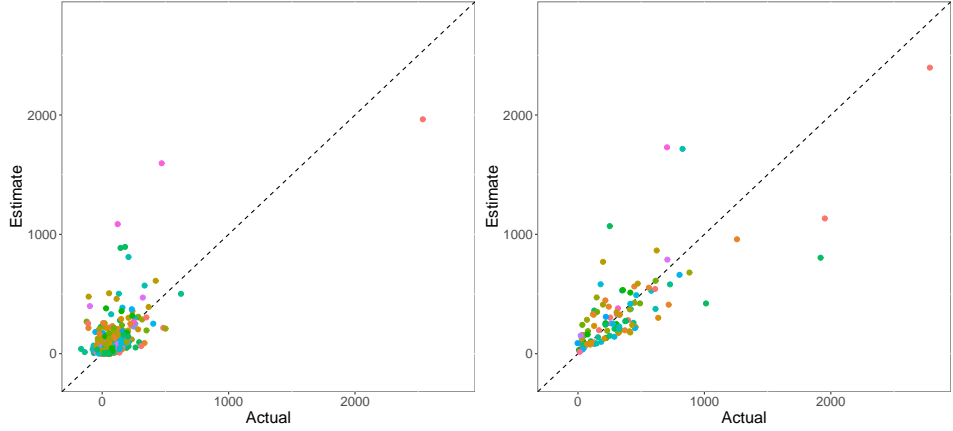}
  \caption{Estimated, $\widetilde{r}_{M}^{T}(\Sigma^{M}-\widetilde{H}_{M})^{-1}\widetilde{r}_{M}$, versus actual, $(n-p)\hat{\sigma}^{2} - (n-p-m)\hat{\sigma}^{2}_{*(M)}$, squared residual differences for the Leave-One-Out (left) and Leave-M-Out (right) cases. Unlike OLS, the estimated and actual values differ because the correlation parameter estimate $\hat{\rho}$ changes. }
\label{fig:GLS} 
\end{figure}  

Figure~\ref{fig:GLS} shows a lack of agreement between actual and estimated SRDs when single observations (left) or all observations for a person (right) are removed from (or added) to the data. Figure~\ref{fig:Rotated} shows the estimated SRDs are less (more) than the actual SRDs when the correlation estimate is smaller (larger) under the reduced dataset model. The relationship between the error in estimation of the SRDs and the change in the correlation parameter estimate is shown in figure~\ref{fig:Linear}. {\em Having both figure 4 and figure 5 is redundant. Maybe show RHS plot of both. Can see that weird slightly off value RHS of figure 5...} 

\begin{figure}
  \includegraphics[width=\linewidth]{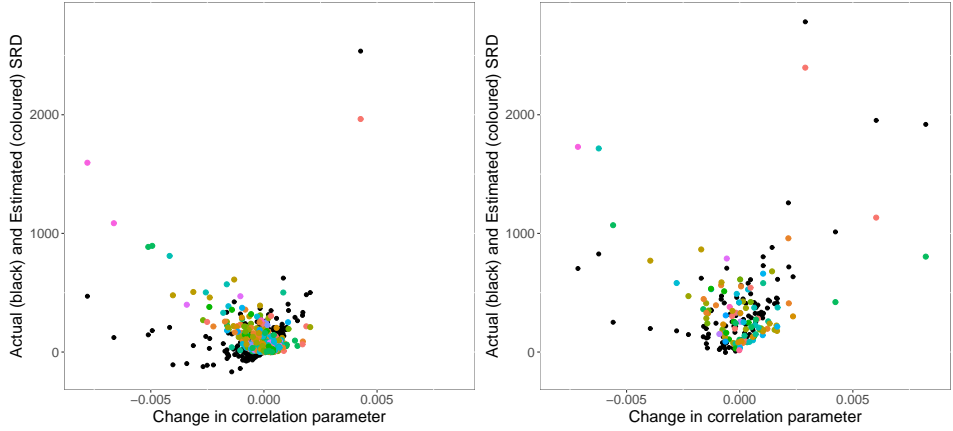}
  \caption{Squared residual differences, actual and estimated, plotted against $\hat{\rho}-\hat{\rho}_{(i)}$, left, or $\hat{\rho}-\hat{\rho}_{(M)}$, right. It is apparent that the estimates are rotated clockwise about $(0,0)$, so that if $\rho$ increases (decreases) under the model fitted to a smaller dataset, $\widetilde{r}_{M}^{T}(\Sigma^{M}-\widetilde{H}_{M})^{-1}\widetilde{r}_{M}$ overestimates (underestimates) $(n-p)\hat{\sigma}^{2} - (n-p-m)\hat{\sigma}^{2}_{*(M)}$.}
\label{fig:Rotated} 
\end{figure} 

\begin{figure}
  \includegraphics[width=\linewidth]{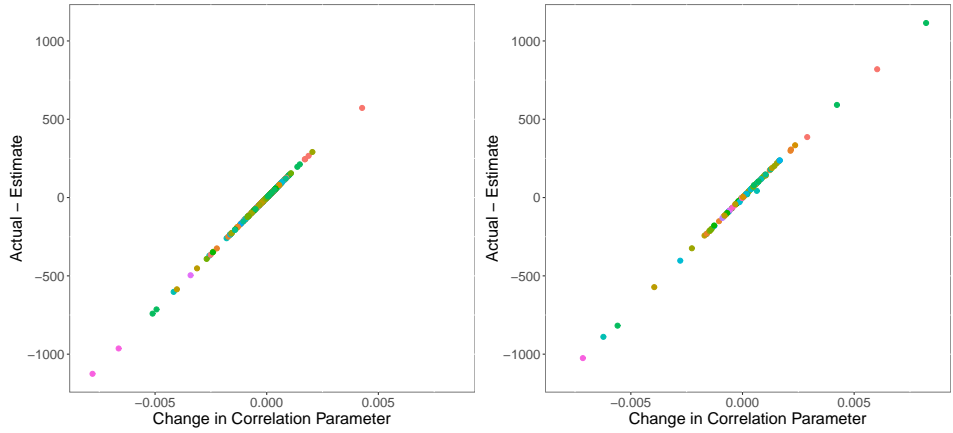}
  \caption{The difference between actual and estimated SRDs is plotted against the change in the correlation estimate $\rho$.}
\label{fig:Linear} 
\end{figure}

If the models fitted to the reduced size datasets force the correlation parameter $\rho$ to be held constant, the plot of actual versus estimated SRDs is like figure 2 rather than figure 3. 

When one observation is left out, for OLS, where the actual changes in the residual sums of squares are very close to the estimated SRDs, the mean actual change in the residual sum of squares is 69.50 and the mean estimated change is 69.44. The mean sum of squares of the leave-one-out cross validation residuals is 71.95. For GLS, where the estimated SRDs are often quite different from the actual SRDs, the actual mean change in the residual sum of squares is 75.10 and the estimated change is 77.28. The mean sum of squares of the leave-one-out cross validation residuals is 78.57.

When the group of observations from each individual is left out in turn, for OLS, the mean actual change in the residual sum of squares is 348.26, close to the mean SRD of 349.22. The mean of the squared leave-M-out cross validation residuals is 379.57. For GLS, the mean actual change in the residual sum of squares is 357.09, surprisingly close to the mean SRD of 357.69. The mean of the squared leave-M-out cross validation residuals is 373.28.   

\begin{figure}
  \includegraphics[width=\linewidth]{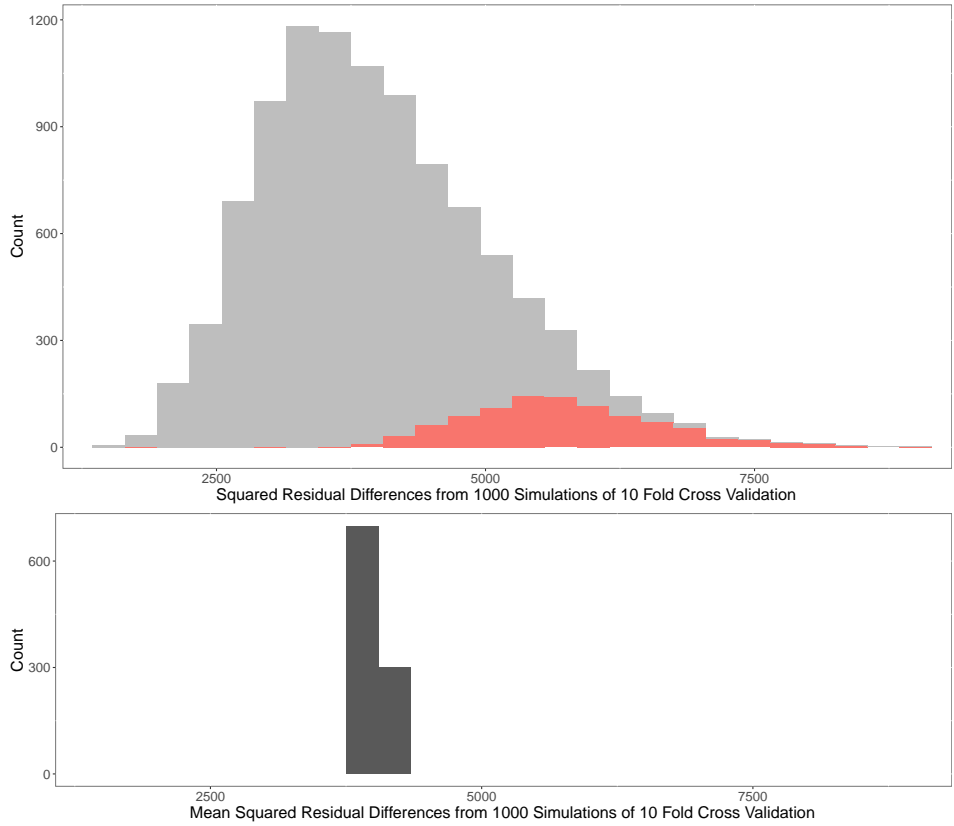}
  \caption{Histograms of SRDs (top) and means of 10 fold SRDs (bottom) for 1000 simulations of 10 fold cross validation on the cervical dystonia dataset. Values from folds which include the unusual observation, number 258, the week 16 value from an individual in the placebo group at site 6, are plotted in red in the top plot. Values from the other 9000 folds are stacked.}
\label{fig:ResidDiffSquared10Fold} 
\end{figure}

For this dystonia example, 1000 simulations of 10 fold cross validation were run. For each simulation, the 522 observations were randomly partitioned in 10 groups, 9 of size $m$ = 52 and the last of size $m$ = 54. The top histogram in figure~\ref{fig:ResidDiffSquared10Fold} shows the estimated SRDs with the values from the folds containing observation 258 stacked in red, below values from the other 9000 folds that did not contain observation 258. The histogram at the bottom of figure~\ref{fig:ResidDiffSquared10Fold} shows the means (over 10 folds) of the 1000 simulations. The individual SRDs have a large range of values but the means do not. 

\section{Conclusion}

The squares of the LMOCV residuals $\widetilde{r}_{M}^{T}(\Sigma^{M}-\widetilde{H}_{M})^{-1}\Sigma^{M}(\Sigma^{M}-\widetilde{H}_{M})^{-1}\widetilde{r}_{M}$ are a reasonable approximation to the squares of the residual differences $\widetilde{r}_{M}^{T}(\Sigma^{M}-\widetilde{H}_{M})^{-1}\widetilde{r}_{M}$ in this example; the Cook's distance quantities are not large since there is not much difference in leverage values between observations.

{\em This paper needs another example - preferable something bigger.}

{\em Need to work out how to roughly approximate the effect of a change in the correlation estimate. ?Look into the code and see how the correlation estimate is arrived at. Maybe show the mean of the SRDs is accurate because correlation averages out to the value from the full dataset.  }

The point of cross validation is to determine if a model will work well on future data ``similar" to existing data, so carrying out a simulation of 10 fold cross validation, where folds are constructed from 50 random observations, is not realistic.  As there were 108 people in the dystonia study, 10 fold cross validation could partition the people, rather than the observations, into folds. None-the-less it is interesting to see the effect of a single unusual observation on entire folds. While cross validation was developed to assess model fit, it also has a use in outlier detection.

\section{Appendix}

We show the derivation of the result for the partitition of the residual sum of squares under the GLS model, where the notation is $\widetilde{Y} = \Sigma^{-1}Y$, $\widetilde{X} = \Sigma^{-1}X$, $\widetilde{r} = \Sigma^{-1}r$ and $\widetilde{H} = \Sigma^{-1}H\Sigma^{-1} = \Sigma^{-1}X(X^{T}\Sigma^{-1}X)^{-1}X^{T}\Sigma^{-1}$. The result for OLS can be derived by substituting $\Sigma=I$, $\widetilde{r} = r$, $\widetilde{H} = X(X^{T}X)^{-1}X^{T}$ and $\Sigma^{M} = I_{m}$. 
\begin{align*}
 & r^{*_{(M)}T}_{(M)}(\Sigma_{(M)})^{-1} r^{*_{(M)}}_{(M)} \\
& \phantom{\hat{\beta}_{(M)}}= (Y_{(M)}-X_{(M)} \hat{\beta}_{(M)})^{T}(\Sigma_{(M)})^{-1}(Y_{(M)}-X_{(M)} \hat{\beta}_{(M)}) \\
& \phantom{\hat{\beta}_{(M)}}= (Y-X\hat{\beta}_{(M)})^{T}\left[ \Sigma^{-1} - \Sigma^{M \textnormal{cols}}(\Sigma^{M})^{-1} \Sigma^{M \textnormal{rows}}\right]
   (Y-X\hat{\beta}_{(M)}) \\
& \phantom{\hat{\beta}_{(M)}}= \begin{multlined}[t](Y-X\hat{\beta}_{(M)})^{T}\Sigma^{-1}(Y-X\hat{\beta}_{(M)}) \\
             - (\widetilde{Y}_{M}-\widetilde{X}_{M}\hat{\beta}_{(M)})^{T}(\Sigma^{M})^{-1}(\widetilde{Y}_{M}-\widetilde{X}_{M}\hat{\beta}_{(M)}) 
   \end{multlined} \\
& \phantom{\hat{\beta}_{(M)}}= \begin{multlined}[t](Y-X\hat{\beta} + X\hat{\beta}-X\hat{\beta}_{(M)})^{T}\Sigma^{-1}(Y-X\hat{\beta}+X\hat{\beta}-X\hat{\beta}_{(M)}) \\ 
             - (\widetilde{Y}_{M}-\widetilde{X}_{M}\hat{\beta}_{(M)})^{T}(\Sigma^{M})^{-1}(\widetilde{Y}_{M}-\widetilde{X}_{M}\hat{\beta}_{(M)}) 
   \end{multlined} \\
& \phantom{\hat{\beta}_{(M)}}= \begin{multlined}[t](Y-X\hat{\beta})^{T}\Sigma^{-1}(Y-X\hat{\beta}) 
             + (X\hat{\beta}-X\hat{\beta}_{(M)})^{T}\Sigma^{-1}(X\hat{\beta}-X\hat{\beta}_{(M)}) \\ 
             - (\widetilde{Y}_{M}-\widetilde{X}_{M}\hat{\beta}_{(M)})^{T}(\Sigma^{M})^{-1}(\widetilde{Y}_{M}-\widetilde{X}_{M}\hat{\beta}_{(M)}) 
   \end{multlined} \\
& \phantom{\hat{\beta}_{(M)}}= \begin{multlined}[t](Y-X\hat{\beta})^{T}\Sigma^{-1}(Y-X\hat{\beta})
             + \widetilde{r}_{M}^{T}(\Sigma^{M}-\widetilde{H}_{M})^{-1}\widetilde{H}_{M}(\Sigma^{M}-\widetilde{H}_{M})^{-1}\widetilde{r}_{M} \\ 
             - \widetilde{r}_{M}^{T}(\Sigma^{M}-\widetilde{H}_{M})^{-1}\Sigma^{M}(\Sigma^{M}-\widetilde{H}_{M})^{-1}\widetilde{r}_{M} \\
   \end{multlined}
\end{align*}

This uses the following results:
\begin{align*}
X_{(M)}^{T}(\Sigma_{(M)})^{-1}X_{(M)} &= X^{T}\Sigma^{-1}X - X^{T}\Sigma^{M \textnormal{cols}}(\Sigma^{M})^{-1} \Sigma^{M \textnormal{rows}}X \\
&= X^{T}\Sigma^{-1}X - \widetilde{X}_{M}^{T}(\Sigma^{M})^{-1} \widetilde{X}_{M}
\end{align*}
The origin of the following matrix inversion result is discussed in \citeauthory{Henderson81}.
\begin{align*} 
& (X^{T}\Sigma^{-1}X - \widetilde{X}_{M}^{T}(\Sigma^{M})^{-1} \widetilde{X}_{M})^{-1} \\
 & \phantom{\hat{\beta}_{(M)}}= (X^{T}\Sigma^{-1}X)^{-1}  + (X^{T}\Sigma^{-1}X)^{-1}\widetilde{X}_{M}^{T}(\Sigma^{M}-\widetilde{H}_{M})^{-1}\widetilde{X}_{M} (X^{T}\Sigma^{-1}X)^{-1}. 
\end{align*}
\begin{align*}
\hat{\beta}_{(M)} &= (X_{(M)}^{T}(\Sigma_{(M)})^{-1}X_{(M)})^{-1}X_{(M)}^{T}(\Sigma_{(M)})^{-1}Y_{(M)} \\
 &= \begin{multlined}[t] \left[(X^{T}\Sigma^{-1}X)^{-1}  
              + (X^{T}\Sigma^{-1}X)^{-1}\widetilde{X}_{M}^{T}(\Sigma^{M}-\widetilde{H}_{M})^{-1}\widetilde{X}_{M} (X^{T}\Sigma^{-1}X)^{-1}\right] \\
                         \times \left[ X^{T}\Sigma^{-1}Y - \widetilde{X}_{M}^{T}(\Sigma^{M})^{-1}\widetilde{Y}_{M} \right] \\
    \end{multlined} \\
 &= \begin{multlined}[t] \hat{\beta} + (X^{T}\Sigma^{-1}X)^{-1}\widetilde{X}_{M}^{T}(\Sigma^{M}-\widetilde{H}_{M})^{-1}\widetilde{X}_{M}\hat{\beta} \\
 - (X^{T}\Sigma^{-1}X)^{-1}\widetilde{X}^{T}_{M}(\Sigma^{M})^{-1}\widetilde{Y}_{M} \\
 - (X^{T}\Sigma^{-1}X)^{-1}\widetilde{X}_{M}^{T}(\Sigma^{M}-\widetilde{H}_{M})^{-1}\widetilde{X}_{M}(X^{T}\Sigma^{-1}X)^{-1}\widetilde{X}^{T}_{M}(\Sigma^{M})^{-1}\widetilde{Y}_{M} 
    \end{multlined} \\
&=  \begin{multlined}[t] \hat{\beta} + (X^{T}\Sigma^{-1}X)^{-1}\widetilde{X}_{M}^{T}(\Sigma^{M} - \widetilde{H}_{M})^{-1} \\
       \times \left[\widetilde{X}_{M}\hat{\beta} - (\Sigma^{M} - \widetilde{H}_{M})(\Sigma^{M})^{-1}\widetilde{Y}_{M} - \widetilde{H}_{M}(\Sigma^{M})^{-1}\widetilde{Y}_{M}\right] 
    \end{multlined} \\
&=  \hat{\beta} - (X^{T}\Sigma^{-1}X)^{-1}\widetilde{X}_{M}^{T}(\Sigma^{M}-\widetilde{H}_{M})^{-1}\widetilde{r}_{M}
\end{align*}
Then
\begin{align*}
\hat{\beta} - \hat{\beta}_{(M)} &= (X^{T}\Sigma^{-1}X)^{-1}\widetilde{X}_{M}^{T}(\Sigma^{M}-\widetilde{H}_{M})^{-1}\widetilde{r}_{M}.
\end{align*}
The next result is Cook's distance for GLS, multiplied by $p \hat{\sigma}^{2}$.
\begin{align*}
&(X\hat{\beta} - X\hat{\beta}_{(M)})^{T}\Sigma^{-1}(X\hat{\beta} - X\hat{\beta}_{(M)}) \\
 & \phantom{\hat{\beta}_{(M)}}= (\hat{\beta} - \hat{\beta}_{(M)})^{T}X^{T}\Sigma^{-1}X(\hat{\beta} - \hat{\beta}_{(M)}) \\
 & \phantom{\hat{\beta}_{(M)}}= \widetilde{r}_{M}^{T}(\Sigma^{M}-\widetilde{H}_{M})^{-1}\widetilde{X}_{M}(X^{T}\Sigma^{-1}X)^{-1}X^{T}\Sigma^{-1}X  \\
 & \phantom{\hat{\beta}_{(M)}\hat{\beta}_{(M)}\hat{\beta}_{(M)}} \times (X^{T}\Sigma^{-1}X)^{-1}\widetilde{X}_{M}^{T}(\Sigma^{M}-\widetilde{H}_{M})^{-1}\widetilde{r}_{M} \\
 & \phantom{\hat{\beta}_{(M)}}= \widetilde{r}_{M}^{T}(\Sigma^{M}-\widetilde{H}_{M})^{-1}\widetilde{H}_{M}(\Sigma^{M}-\widetilde{H}_{M})^{-1}\widetilde{r}_{M} 
\end{align*}
\begin{align*}
\widetilde{Y}_{M}-\widetilde{X}_{M}\hat{\beta}_{(M)} 
 &= \widetilde{Y}_{M}-\widetilde{X}_{M}\hat{\beta}+\widetilde{X}_{M}\hat{\beta}-\widetilde{X}_{M}\hat{\beta}_{(M)} \\
 &= \widetilde{r}_{M} + \widetilde{X}_{M}(X^{T}\Sigma^{-1}X)^{-1}\widetilde{X}_{M}^{T}(\Sigma^{M} - \widetilde{H}_{M})^{-1}\widetilde{r}_{M} \\
 &= \widetilde{r}_{M} + \widetilde{H}_{M}(\Sigma^{M} - \widetilde{H}_{M})^{-1}\widetilde{r}_{M} \\
 &= \Sigma^{M}(\Sigma^{M} - \widetilde{H}_{M})^{-1}\widetilde{r}_{M}
\end{align*}
For known $\Sigma$,
\begin{align*}
(Y-X\hat{\beta})^{T}\Sigma^{-1}(X\hat{\beta}-X\hat{\beta}_{(M)}) &= (X\hat{\beta}-X\hat{\beta}_{(M)})^{T}\Sigma^{-1} (Y-X\hat{\beta}) \\ 
 &= 0 
\end{align*}

\end{document}